\documentclass{webofc}

\usepackage[varg]{txfonts}   
\usepackage{hyperref}
\usepackage{url}
\hypersetup{colorlinks=true,citecolor=blue,urlcolor=blue,linkcolor=blue}
\newcommand{\Pom}{\mathbb{P}}
\newcommand{\Reg}{\mathbb{R}}
\newcommand{\bpta}{\mbox{\boldmath $p_{t,1}$}}
\newcommand{\bptb}{\mbox{\boldmath $p_{t,2}$}}
\newcommand{\bkt}{\mbox{\boldmath $k_{t}$}}
\newcommand{\bktsqrt}{\mbox{\boldmath{$k_{t}^{2}$}}}
\newcommand{\bptat}{\mbox{\boldmath $\tilde{p}_{t,1}$}}
\newcommand{\bptbt}{\mbox{\boldmath $\tilde{p}_{t,2}$}}
\renewcommand\slash[1]{\not \!\! #1}
\begin{document}
\title{Central exclusive production of $\eta$ and $\eta'$ mesons \\in diffractive proton-proton collisions at the LHC}
%
%

\author{\firstname{Piotr} \lastname{Lebiedowicz}\inst{1}\fnsep\thanks{\email{Piotr.Lebiedowicz@ifj.edu.pl}} \and
        \firstname{Otto} \lastname{Nachtmann}\inst{2}\fnsep\thanks{\email{O.Nachtmann@thphys.uni-heidelberg.de}} \and
        \firstname{Antoni} \lastname{Szczurek}\inst{1, 3}\fnsep\thanks{\email{Antoni.Szczurek@ifj.edu.pl}}
}

\institute{Institute of Nuclear Physics Polish Academy of Sciences, \\Radzikowskiego 152, PL-31342 Krak\'ow, Poland
\and
           Institut f\"ur Theoretische Physik, Universit\"at Heidelberg, \\Philosophenweg 16, D-69120 Heidelberg, Germany
\and
           Institute of Physics, Faculty of Exact and Technical Sciences, University of Rzesz\'ow,\\
Pigonia 1, PL-35310 Rzesz\'ow, Poland
          }

\abstract{We discuss central exclusive production (CEP) of $\eta$ and $\eta'(958)$ mesons in high-energy proton-proton collisions using the tensor-pomeron approach, incorporating absorption effects at the amplitude level. Model parameters are constrained using WA102 experimental data measured at $\sqrt{s} = 29.1$ GeV. We present cross sections and differential distributions for the LHC energy of $\sqrt{s} = 13$ TeV. For $pp \to pp \eta$, the upper limit of the total cross section is 2.5 $\mu$b for $|\eta_{M}| < 1$ and 5.6 $\mu$b for $2 < \eta_{M} < 5$. For $pp \to pp \eta'$, the predicted cross sections are 0.3--0.7 $\mu$b and 0.9--2.1 $\mu$b in the respective pseudorapidity regions. These results demonstrate the feasibility of studying diffractive flavour singlet pseudoscalar meson production to probe the nature of the pomeron at the LHC.}


%
\maketitle
\section{Introduction}
\label{intro}

\begin{figure}[h]
\centering
\includegraphics[width=5cm,clip]{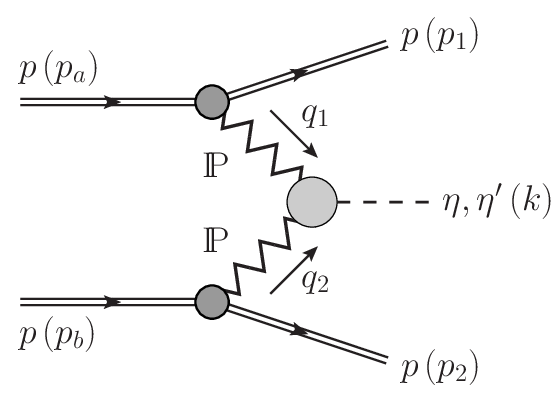}
\caption{CEP of $\eta$ and $\eta'(958)$ with double-pomeron exchange.
The pomeron momenta are $q_{1}$, $q_{2}$, and $k = q_{1} + q_{2}$
is the momentum of the produced meson. We have $m_{M}^{2} = k^{2}$
($M = \eta$ or $\eta'(958)$).}
\label{fig:1}
\end{figure}

In this contribution we discuss central exclusive production (CEP) of pseudoscalar ($J^{PC} = 0^{-+}$) mesons
in proton-proton collisions,
\begin{eqnarray}
p + p \to p +  M + p \,, \quad M = \eta \; {\rm or} \; \eta'(958)\,.
\label{1.1}
\end{eqnarray}
We discuss these reactions for the c.m. energy $\sqrt{s} = 29.1$ GeV
realised at the WA102 experiment \cite{Kirk:2000ws,WA102:1998ixr}
and for $\sqrt{s} = 13$ TeV corresponding to the LHC experiments.
We require large rapidity gaps between $M$ and the outgoing protons.
At high energies the pomeron-pomeron ($\Pom-\Pom$) fusion process
(figure~\ref{fig:1}) is expected to be dominant. 
For a review see \cite{Donnachie:2002en}.

The presentation is based on \cite{Lebiedowicz:2025num}
where all details and many more results can be found.
We improve the results given in \cite{Lebiedowicz:2013ika} 
and extend them to LHC energies.
We treat the $pp \to pp \eta(\eta')$ reactions 
within the tensor-pomeron approach \cite{Ewerz:2013kda}.
The secondary exchanges ($f_{2 \Reg}-\Pom$, 
$\Pom-f_{2 \Reg}$, and $f_{2 \Reg}-f_{2 \Reg}$) 
give small contributions at the LHC energies, at least,
in the midrapidity region.
The secondary reggeons play an important role in the WA102 energy range,
see e.g. \cite{Lebiedowicz:2013ika}
and the discussion in \cite{Lebiedowicz:2020yre} for the CEP of $f_{1}$ mesons.

Some remarks on different views of the pomeron were made 
in \cite{Ewerz:2016onn}. 
In \cite{Ewerz:2016onn} it was shown that a scalar pomeron
is not compatible with the measurements by STAR \cite{Adamczyk:2012kn}
concerning the helicity structure of $pp$ elastic scattering.

Exclusive production of $f_{1}(1285)$ and $\eta$ mesons
in proton-proton collisions at energies available at the GSI-FAIR
is discussed in \cite{Lebiedowicz:2021gub,Lebiedowicz:2025nhh}.
At energies close to the threshold the $VV$-fusion ($V = \rho, \omega$) mechanism,
and the mechanism involving nucleon resonances ($N^{*}$) excited 
via the pseudoscalar- and/or vector-meson exchanges play a dominant role.

\section{Brief overview of the formalism}
\label{sec-1}

The amplitude for the reaction (\ref{1.1}) reads
\begin{eqnarray}
{\cal {M}}_{pp \to pp M} =
{\cal {M}}_{pp \to pp M}^{\rm Born} + 
{\cal {M}}_{pp \to pp M}^{pp-\rm{rescattering}}\,.
\label{amp_PP}
\end{eqnarray}
The second term represents the $pp$-rescattering corrections
to the first (Born) term.
The Born amplitude via pomeron-pomeron fusion is given by
\begin{equation}
\begin{split}
{\cal M}^{\rm Born}_{\lambda_{a}\lambda_{b}\to\lambda_{1}\lambda_{2} M} 
= & (-i)\,
\bar{u}(p_{1}, \lambda_{1}) 
i\Gamma^{(\Pom pp)\,\mu_{1} \nu_{1}}(p_{1},p_{a}) 
u(p_{a}, \lambda_{a})\\
& \times 
i\Delta^{(\Pom)}_{\mu_{1} \nu_{1}, \alpha_{1} \beta_{1}}(2 \nu_{1},t_{1})\,
i\Gamma^{(\Pom \Pom M)\,\alpha_{1} \beta_{1},\alpha_{2} \beta_{2}}(q_{1},q_{2})\,
i\Delta^{(\Pom)}_{\alpha_{2} \beta_{2}, \mu_{2} \nu_{2}}(2 \nu_{2},t_{2}) \\
& \times
\bar{u}(p_{2}, \lambda_{2}) 
i\Gamma^{(\Pom pp)\,\mu_{2} \nu_{2}}(p_{2},p_{b}) 
u(p_{b}, \lambda_{b}) \,,
\end{split}
\label{amplitude}
\end{equation}
where $p_{a,b}$, $p_{1,2}$ and $\lambda_{a,b}$, 
$\lambda_{1,2} = \pm \frac{1}{2}$
denote the four-momenta and helicities of the protons, respectively.
The relevant kinematic quantities are
\begin{eqnarray}
&&s = (p_{a} + p_{b})^{2}, \quad
s_{1} = (p_{a} + q_{2})^{2} = (p_{1} + k)^{2}, \quad
s_{2} = (p_{b} + q_{1})^{2} = (p_{2} + k)^{2},
\nonumber\\
&&
k = q_{1} + q_{2}, \quad
q_1 = p_{a} - p_{1}, \quad
q_2 = p_{b} - p_{2}, \quad
t_1 = q_{1}^{2}, \quad
t_2 = q_{2}^{2}, \nonumber\\
&&
u_1 = (p_{a} - k)^{2}, \quad
u_2 = (p_{b} - k)^{2}, \quad
\nu_{1} = \frac{1}{4}(s_{1} - u_{1}), \quad
\nu_{2} = \frac{1}{4}(s_{2} - u_{2})\,.
\label{1.2}
\end{eqnarray}
The expressions
for the effective pomeron-proton vertex function
and propagator for the tensor-pomeron exchange
are (see Sect.~3 of \cite{Ewerz:2013kda})
\begin{eqnarray}
&&i\Gamma_{\mu \nu}^{(\Pom pp)}(p',p)
=-i 3 \beta_{\Pom NN} F_{1}(t)
\left(
\frac{1}{2} \gamma_{\mu}(p'+p)_{\nu} 
+ \frac{1}{2} \gamma_{\nu}(p'+p)_{\mu}
- \frac{1}{4} g_{\mu \nu} (\slash{p}' + \slash{p})
\right), \quad \;
\label{add2}\\
&&i \Delta^{(\Pom)}_{\mu \nu, \kappa \lambda}(2 \nu_{1,2},t) = 
\frac{1}{8\nu_{1,2}} 
\left( g_{\mu \kappa} g_{\nu \lambda} 
     + g_{\mu \lambda} g_{\nu \kappa}
     - \frac{1}{2} g_{\mu \nu} g_{\kappa \lambda} \right)
(-i \,2 \nu_{1,2} \,\alpha'_{\Pom})^{\alpha_{\Pom}(t)-1}\,,
\label{add1}
\end{eqnarray}
where $t = (p'-p)^{2}$ and $\beta_{\Pom NN} = 1.87$~GeV$^{-1}$.
For simplicity we use for the pomeron-proton coupling 
the electromagnetic Dirac form factor $F_{1}(t)$ of the proton.
We use $2 \nu_{1}$ and $2 \nu_{2}$
instead of $s_{1}$ and $s_{2}$, respectively,
as energy variables in the pomeron propagators.
For large $s_{1,2}$ and small $|t_{1,2}|$ we have $2 \nu_{1,2} \to s_{1,2}$.
The pomeron trajectory $\alpha_{\Pom}(t)$ is
assumed to have the standard form:
\begin{eqnarray}
\alpha_{\Pom}(t) = \alpha_{\Pom}(0)+\alpha'_{\Pom} t\,,
\quad 
\alpha_{\Pom}(0) = 1 + \epsilon_{\Pom} = 1.0808\,, 
\quad
\alpha'_{\Pom} = 0.25 \; \mathrm{GeV}^{-2}\,.
  \label{pomtrajectory}
\end{eqnarray}

The $\Pom \Pom M$ coupling was discussed 
in Sect.~2.2 of \cite{Lebiedowicz:2013ika} 
and in \cite{Lebiedowicz:2025num}.
The resulting $\Pom \Pom M$ vertex, including a form factor, 
is given as follows
\begin{eqnarray}
&&i\Gamma_{\mu \nu,\kappa \lambda}^{(\Pom \Pom M)} (q_{1},q_{2}) =
\left( 
  i\Gamma_{\mu \nu,\kappa \lambda}'^{(\Pom \Pom M)}(q_{1},q_{2}) \mid_{\rm bare}
+ i\Gamma_{\mu \nu,\kappa \lambda}''^{(\Pom \Pom M)}(q_{1},q_{2}) \mid_{\rm bare} 
\right)
F(q_{1}^{2},q_{2}^{2})\,,
\label{vertex_pompomPS}\\
&&i\Gamma_{\mu \nu,\kappa \lambda}'^{(\Pom \Pom M)}(q_{1}, q_{2})\mid_{\rm bare} =
i \, \frac{g_{\Pom \Pom M}'}{2 M_{0}} \,
\left( g_{\mu \kappa} \varepsilon_{\nu \lambda \rho \sigma}
      +g_{\nu \kappa} \varepsilon_{\mu \lambda \rho \sigma}
      +g_{\mu \lambda}\varepsilon_{\nu \kappa \rho \sigma}
      +g_{\nu \lambda}\varepsilon_{\mu \kappa \rho \sigma} \right) 
\nonumber \\
&&\qquad \qquad \qquad \qquad \quad \times \;
(q_{1}-q_{2})^{\rho} (q_{1}+q_{2})^{\sigma}\,,
\label{2.1}\\
&&i\Gamma_{\mu \nu,\kappa \lambda}''^{(\Pom \Pom M)}(q_{1}, q_{2})\mid_{\rm bare} =
i \, \frac{g_{\Pom \Pom M}''}{M_{0}^{3}} \, 
\bigg{\lbrace} \varepsilon_{\nu \lambda \rho \sigma} \left[ q_{1 \kappa} q_{2 \mu} - (q_{1} \cdot q_{2}) g_{\mu \kappa} \right]  \nonumber \\
&&\qquad \qquad \qquad \qquad \quad +
\varepsilon_{\mu \lambda \rho \sigma} \left[ q_{1 \kappa} q_{2 \nu} - (q_{1} \cdot q_{2}) g_{\nu \kappa} \right] + 
\varepsilon_{\nu \kappa \rho \sigma}  \left[ q_{1 \lambda} q_{2 \mu} - (q_{1} \cdot q_{2}) g_{\mu \lambda} \right] \nonumber \\
&&\qquad \qquad \qquad \qquad \quad +
\varepsilon_{\mu \kappa \rho \sigma}  \left[ q_{1 \lambda} q_{2 \nu} - (q_{1} \cdot q_{2}) g_{\nu \lambda} \right] 
\bigg{\rbrace}
(q_{1}-q_{2})^{\rho} (q_{1}+q_{2})^{\sigma}\,. \qquad
\label{2.2}
\end{eqnarray}
Here
$M_{0} \equiv 1$~GeV
and $g'_{\Pom \Pom M}$, $g''_{\Pom \Pom M}$ are dimensionless coupling constants that should be fitted to experimental data
(see table~I of \cite{Lebiedowicz:2025num} for the coupling constants used in our calculations, obtained from fits to the WA102 data).
The $\Gamma_{\mu \nu,\kappa \lambda}'^{(\Pom \Pom M)}$ and 
$\Gamma_{\mu \nu,\kappa \lambda}''^{(\Pom \Pom M)}$ bare
vertices correspond to
$(l,S) = (1,1)$ and $(3,3)$, respectively,
as derived from the corresponding coupling Lagrangians.
The form factor in the central vertex is parametrised as
\begin{eqnarray}
F(t_{1},t_{2}) = 
\exp\left( \frac{t_{1}+t_{2}}{\Lambda_{E}^{2}}\right) \,,
\label{Fpompommeson_exp}
\end{eqnarray}
where the cutoff parameter $\Lambda_{E} \simeq 1.0\;{\rm GeV}$ is adjusted 
to experimental data; see table~I of \cite{Lebiedowicz:2025num}.

The amplitude representing absorption corrections 
due to the proton-proton interactions
can be written as
\begin{eqnarray}
{\cal M}_{pp \to pp M}^{pp-\rm{rescattering}}(s,\bpta,\bptb)=
\frac{i}{8 \pi^{2} s} \int d^{2}\bkt \,
{\cal M}_{pp\to pp M}^{\rm Born}(s,\bptat,\bptbt)
{\cal M}_{pp \to pp}(s,t)\,,
\label{abs_correction}
\end{eqnarray}
where $\bpta$ and $\bptb$
are the transverse components of the momenta of the outgoing protons
and $\bkt$
is the transverse momentum in the pomeron loop.
${\cal M}_{pp\to pp M}^{\rm Born}$
is the Born amplitude given by (\ref{amplitude})
with $\bptat = \bpta - \bkt$ and $\bptbt = \bptb + \bkt$.
${\cal M}_{pp \to pp}$
is the elastic $pp$ scattering amplitude
for large $s$ and with the momentum transfer $t=-\bktsqrt$.
We work with the amplitudes in the high-energy approximation,
where $s$-channel-helicity conservation 
of the protons holds.

In addition to the $\Pom \Pom$ fusion, 
we also consider the $\Pom f_{2 \Reg}$, $f_{2 \Reg} \Pom$,
and $f_{2 \Reg} f_{2 \Reg}$ exchanges.
Since $\Pom$ and $f_{2 \Reg}$ are described as $(C = +1)$ tensor exchanges, 
the additional amplitudes, 
${\cal {M}}^{(\Pom f_{2 \Reg})}$, ${\cal {M}}^{(f_{2 \Reg} \Pom)}$, and ${\cal {M}}^{(f_{2 \Reg} f_{2 \Reg})}$,
are treated similarly to ${\cal {M}}^{(\Pom \Pom)}$.
The effective $f_{2 \Reg}$-proton vertex function
and the $f_{2 \Reg}$ propagator are similar to $\Pom$.
The important differences there are 
in the magnitude of the coupling constant 
and in the parametrisation of the trajectory:
\begin{eqnarray}
&&3 \beta_{\Pom NN} \to \frac{g_{f_{2 \Reg} pp}}{M_{0}}\,, \quad g_{f_{2 \Reg} pp} = 11.04\,,\nonumber \\
&&\alpha_{\Pom}(t) \to \alpha_{\Reg_{+}}(t) =
\alpha_{\Reg_{+}}(0)+\alpha'_{\Reg_{+}}t\,, \quad
\alpha_{\Reg_{+}}(0) = 0.5475\,, \quad
\alpha'_{\Reg_{+}} = 0.9 \; \mathrm{GeV}^{-2}\,.
\label{amp_PR}
\end{eqnarray}

\section{Results}
\label{sec-2}

\subsection{Comparison with the WA102 data}
\label{sec-3}

\begin{figure}[h]
\centering
\includegraphics[width=6cm,clip]{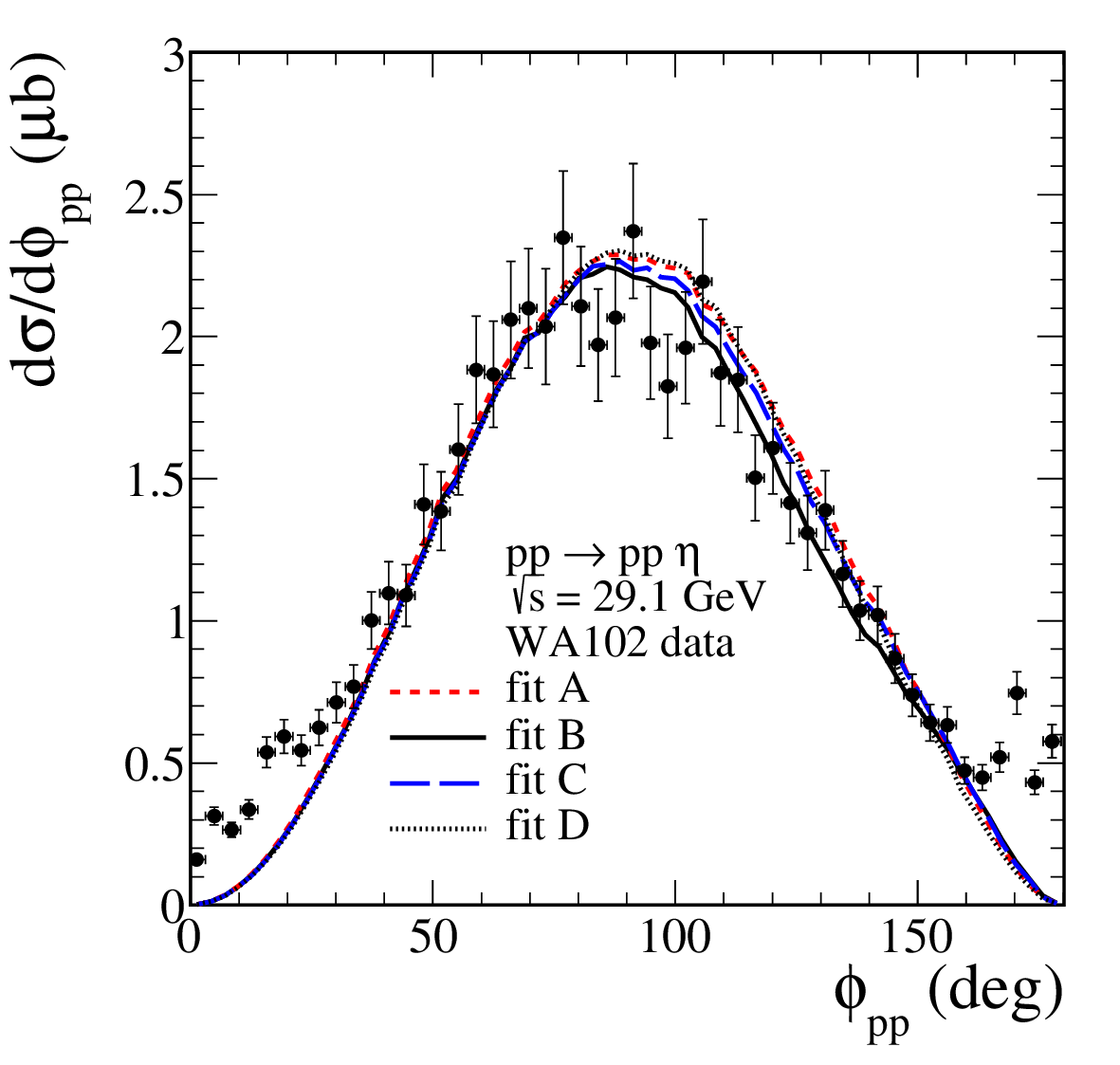}
\includegraphics[width=6cm,clip]{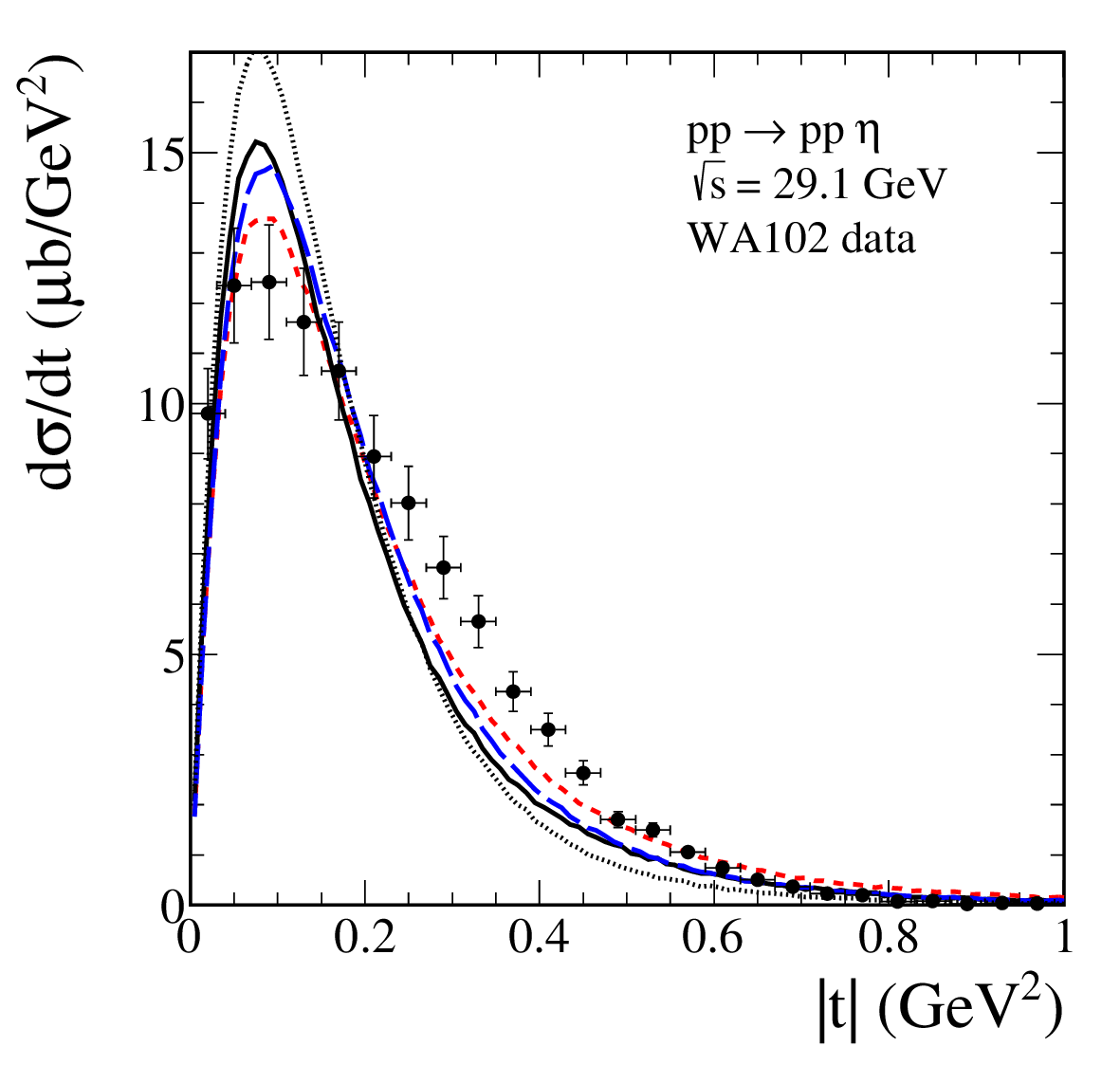}
\caption{Data for CEP of $\eta$ mesons [$M = \eta$ in (\ref{1.1})]
from the WA102 experiment
compared to fits from the tensor-pomeron model.
The WA102 data points from \cite{WA102:1998ixr}
have been normalised to the mean value of the total cross section
(\ref{3.1}).
Left panel:
Distribution of $\phi_{pp}$, the angle between the transverse
momenta of the outgoing protons in (\ref{1.1}), see figure~\ref{fig:1}.
Right panel:
Distribution in $t = (p_{a} - p_{1})^{2}$ or $(p_{b} - p_{2})^{2}$ 
in (\ref{1.1}); $t = t_{1}$ or $t_{2}$ see (\ref{1.2}). 
(From figure~3 of \cite{Lebiedowicz:2025num}).}
\label{fig:2}
\end{figure}

\begin{figure}[h]
\centering
\includegraphics[width=6cm,clip]{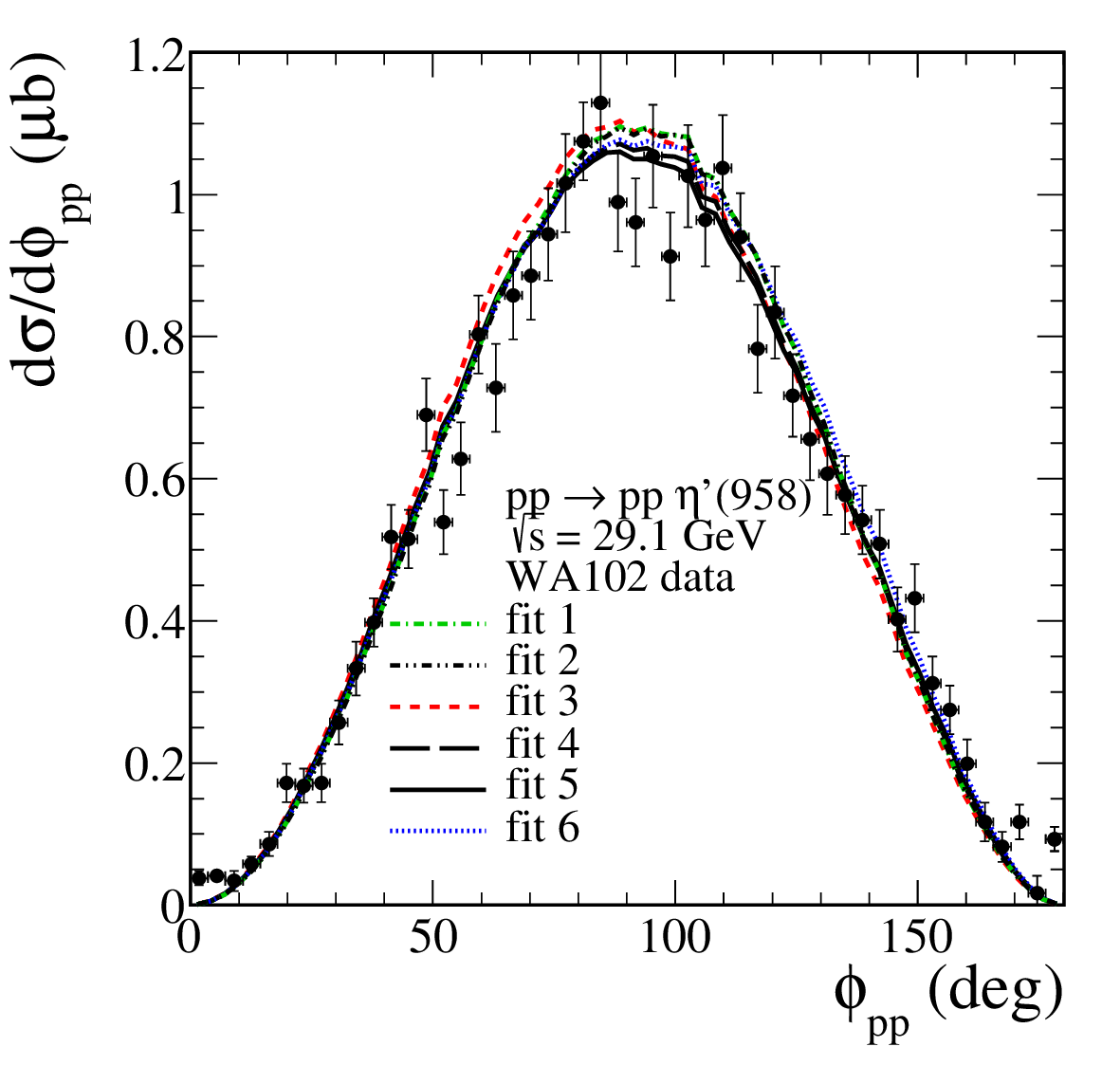}
\includegraphics[width=6cm,clip]{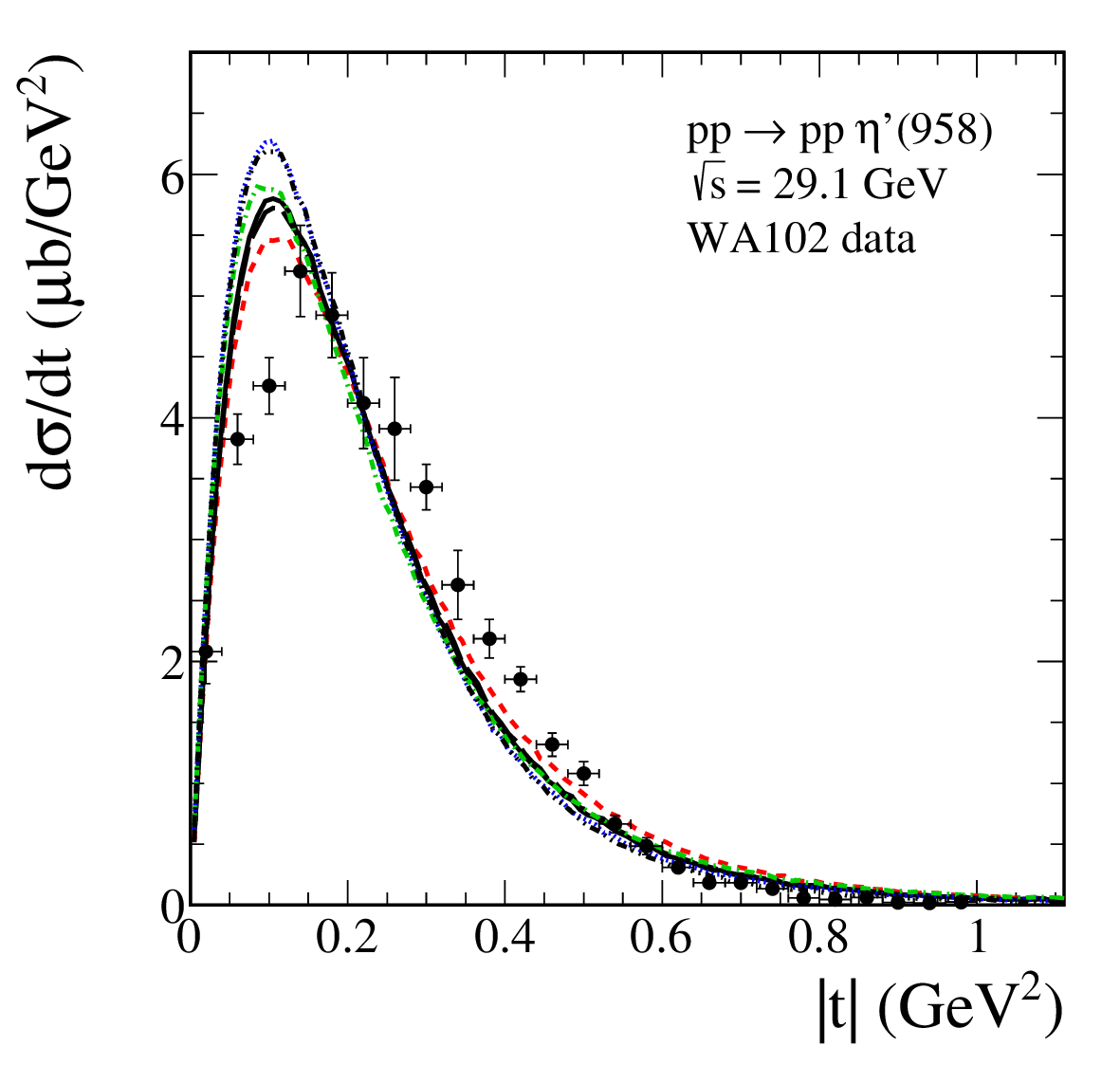}
\caption{Same as figure~\ref{fig:2} but for $\eta'$ CEP.
(From figure~2 of \cite{Lebiedowicz:2025num}).}
\label{fig:3}
\end{figure}

There exist measurements of $\eta$ and $\eta'(958)$ CEP 
from the WA102 experiment at c.m. energy $\sqrt{s} = 29.1$~GeV.
The total cross sections are \cite{Kirk:2000ws}:
\begin{eqnarray}
\eta:  \sigma_{\rm{exp.}} = (3859 \pm 368)~{\rm nb}\,, \qquad
\eta': \sigma_{\rm{exp.}} = (1717 \pm 184)~{\rm nb}\,.
\label{3.1}
\end{eqnarray}
In the WA102 experiment many differential distributions were measured.
As examples we show in figure~\ref{fig:2} for $\eta$
and in figure~\ref{fig:3} for $\eta'$ CEP
the distributions in $\phi_{pp}$,
the angle between the transverse momenta of the outgoing protons
$p(p_{1})$ and $p(p_{2})$,
and the $t$ distributions.
Here, $t$ is the four-momentum transfer squared 
from one of the proton vertices
(we have $t = t_{1}$ or $t_{2}$).
In order to describe the $\phi_{pp}$ distributions
with the maximum at $\phi_{pp} \simeq \pi/2$ we need
both couplings $g'$ of (\ref{2.1}) and $g''$ of (\ref{2.2}).
The dominant contribution to the cross section 
comes from the coupling $(l,S) = (1,1)$ (\ref{2.1}).

The relative contributions of pomeron
and non-leading exchanges are varied across various fit scenarios.
For the description of $\eta$ production 
the $\Pom \Pom$, $\Pom f_{2 \Reg}$, $f_{2 \Reg} \Pom$,
and $f_{2 \Reg} f_{2 \Reg}$
exchanges in the amplitude are included.
In the calculations, we use four sets of parameters 
corresponding to Fits A--D.
In Set A, we assume that the $\Pom \Pom$-exchange contribution 
is more important than in the other sets.
We calculate the percentage share of different exchanges to the total cross section at $\sqrt{s} = 29.1$~GeV.
In Fit~C, $\Pom \Pom$ represents 40\%, 
$\Pom f_{2 \Reg}$ + $f_{2 \Reg} \Pom$ accounts for 32\%,
and $f_{2 \Reg} f_{2 \Reg}$ for 5\%.
In Fit~D we have 25\%, 48\%, and 4\%, respectively.
For the $pp \to pp \eta'$ reaction,
in Fits 1--3, 
we consider $\Pom \Pom$ plus $f_{2 \Reg} f_{2 \Reg}$ fusion processes,
while Fits~4--6 extend the calculation by including the
$\Pom f_{2 \Reg}$ plus $f_{2 \Reg} \Pom$ exchanges.
We calculate the percentage share of different exchanges to the total cross section at $\sqrt{s} = 29.1$~GeV.
In Fit~5, $\Pom \Pom$ accounts for 39\%, 
$\Pom f_{2 \Reg}$ + $f_{2 \Reg} \Pom$ for 28\%,
and $f_{2 \Reg} f_{2 \Reg}$ for only 3\%.
There is a large interference effect between these terms 
in the amplitude of about 30\% with respect to the total cross section.
We found that the WA102 energy
is too low for pinning down the pomeron contribution.

\subsection{Predictions for the LHC experiments}
\label{sec-4}

\begin{figure}[h]
\centering
\includegraphics[width=6cm,clip]{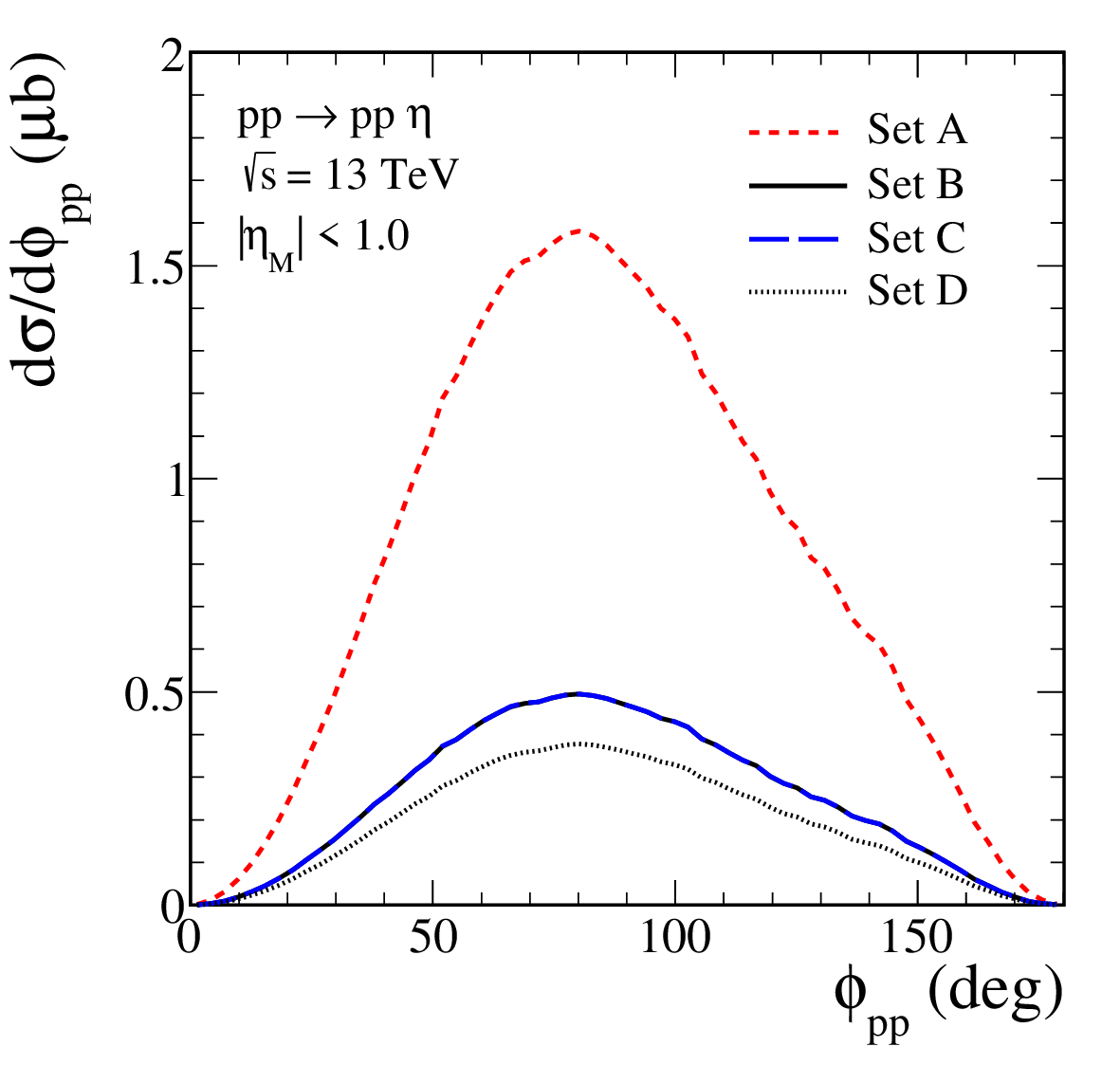}
\includegraphics[width=6cm,clip]{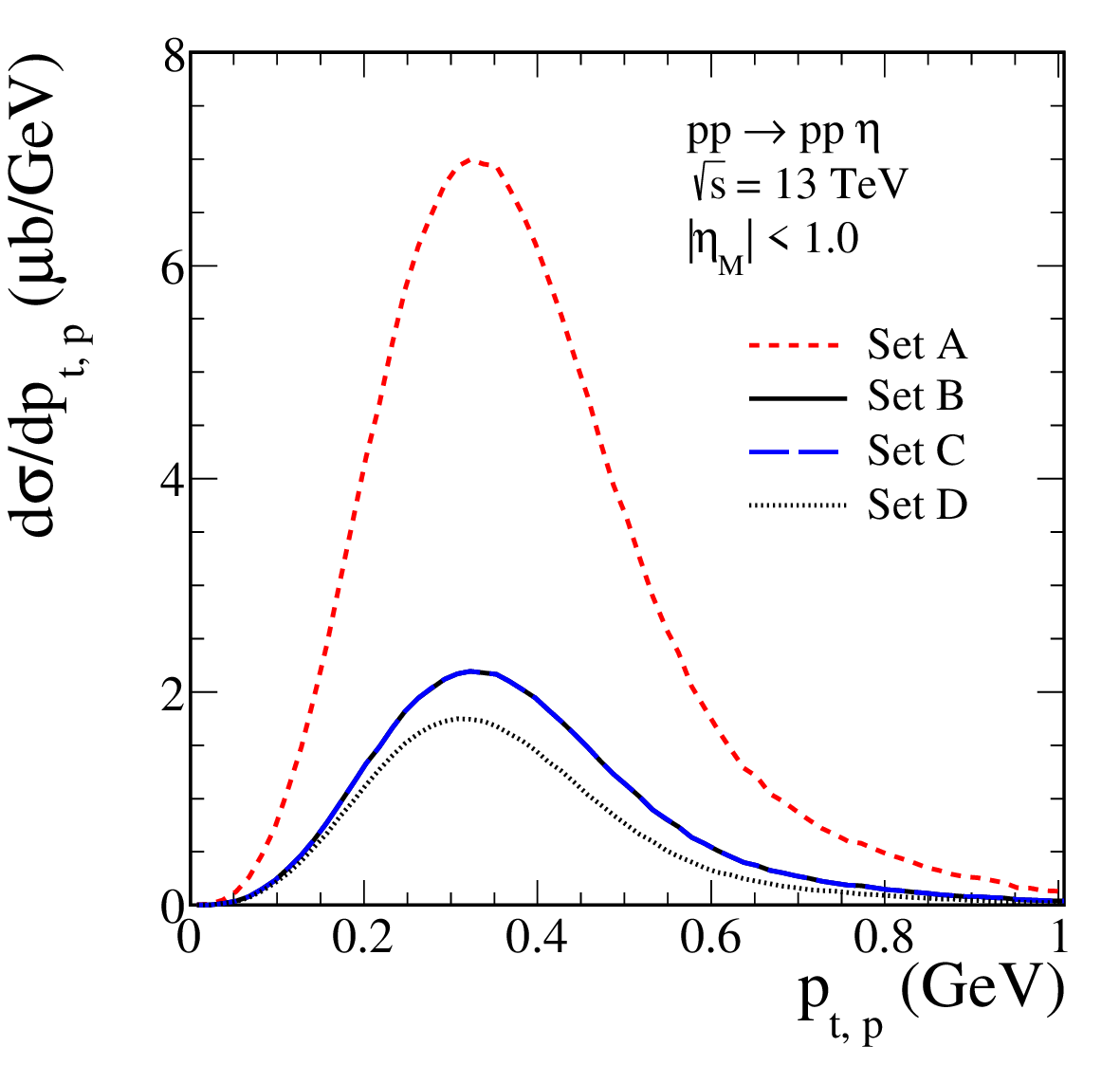}
\caption{Predictions for CEP of the $\eta$ meson 
calculated at c.m. energy $\sqrt{s} = 13$~TeV 
and for $|\eta_{M}| < 1.0$:
$\phi_{pp}$ distributions and $p_{t,p}$ distributions.
(From figure~7 of \cite{Lebiedowicz:2025num}).}
\label{fig:4}
\end{figure}

\begin{figure}[h]
\centering
\includegraphics[width=6cm,clip]{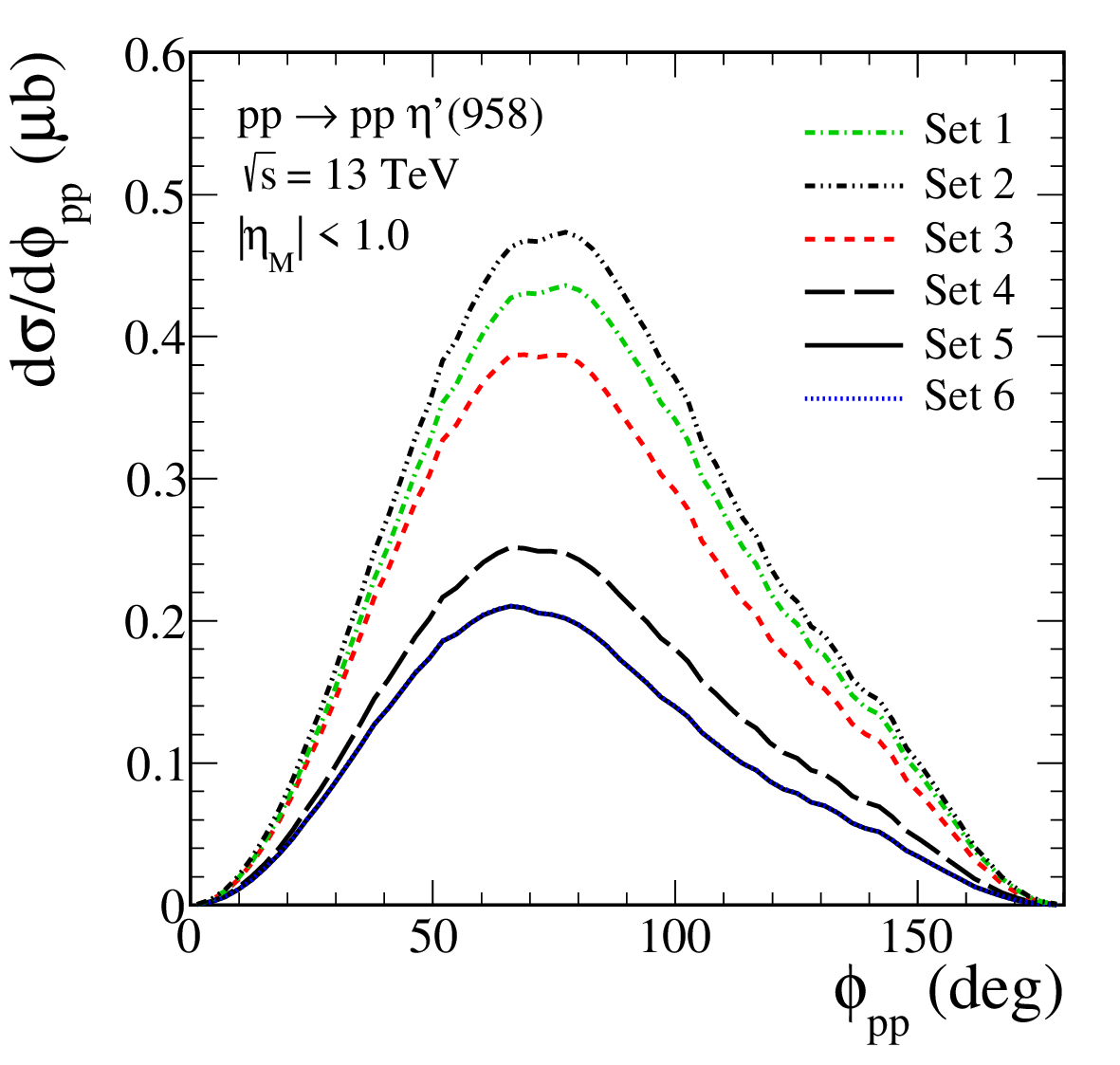}
\includegraphics[width=6cm,clip]{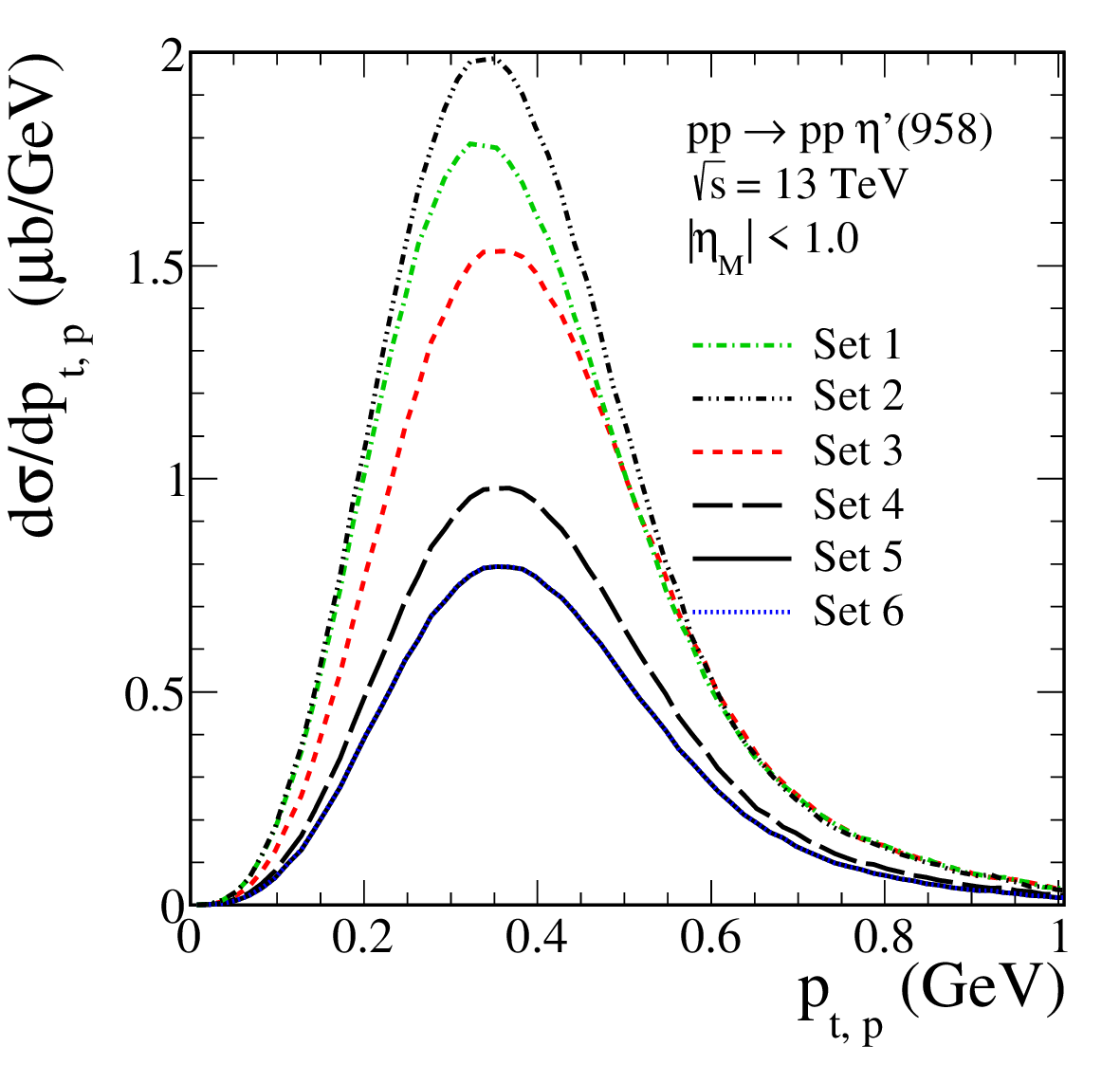}
\caption{Same as figure~\ref{fig:2} but for $\eta'$ CEP.
(From figure~5 of \cite{Lebiedowicz:2025num}).}
\label{fig:5}
\end{figure}

Now we present our predictions for the LHC experiments
where subleading $f_{2 \Reg}$ contributions should be very small.
We choose here $\sqrt{s} = 13$~TeV.
In figures~\ref{fig:4} and \ref{fig:5} we show our predictions
for $\eta$ and $\eta'$ CEP where we use the following variables:
$\phi_{pp}$ angle between the transverse momenta of the outgoing protons
$p(p_{1})$ and $p(p_{2})$,
$p_{t,p}$ absolute value of the transverse momentum of the protons
$p(p_{1})$ or $p(p_{2})$.

Our fits which were practically indistinguishable at the WA102 energy 
($\sqrt{s} = 29.1$ GeV) 
give quite different predictions
for $\sqrt{s}=13$~TeV.
Our predictions for CEP of $\eta$ and $\eta'$
based on parameter set~A and sets 1--3, respectively,
(obtained assuming significant double-pomeron exchange 
at the WA102 energy) 
should be considered as upper limits of the cross sections.
For sets~4--6, the cross section is smaller by a factor of about~2;
see table~II of \cite{Lebiedowicz:2025num}.
At the LHC, absorption effects play a major role, 
significantly reducing the cross sections.
For $pp \to pp \eta$, we predict an upper limit
for the total cross section 
of 2.5~$\mu$b for pseudorapidity of the $\eta$ meson $|\eta_{M}| < 1$
and 5.6~$\mu$b for $2 < \eta_{M} < 5$.
For $pp \to pp \eta'$, we predict the cross section
to be in the range of
0.3--0.7~$\mu$b for pseudorapidity of the $\eta'$ meson $|\eta_{M}| < 1$
and 0.9--2.1~$\mu$b for $2 < \eta_{M} < 5$.

\section{Conclusions}
\label{sec-5}

We have discussed CEP of $\eta$ and $\eta'$ mesons 
in proton-proton collisions at high energies.
Data exist only from the WA102 experiment and, hopefully, 
will be collected in the future by LHC experiments.
Our main points are as follows.

\begin{itemize}
\item
We conclude that since the fusion of two scalar 
pomerons cannot produce the $J^{PC} = 0^{-+}$ quantum 
numbers of the pseudoscalar mesons, the 
tensor-pomeron approach offers a clear advantage in 
correctly describing the coupling structure of these 
diffractive vertices. Consequently, the experimental 
observation of CEP of $\eta$, $\eta'(958)$, or $f_{1}(1285)$ mesons 
at the LHC would be strictly incompatible with a scalar character 
of the pomeron. In contrast, the tensor-pomeron 
model \cite{Ewerz:2013kda,Lebiedowicz:2025num} naturally accounts for 
the production of these states via the $\Pom \Pom$ 
fusion, making their measurement a good
test of the pomeron couplings.

We emphasize that a vector-like pomeron coupling 
cannot be applied, as it violates the charge-conjugation 
($C = +1$) property required for pomeron-exchange \cite{Ewerz:2013kda,Ewerz:2016onn}. 
Furthermore, it incorrectly predicts opposite signs for 
$pp$ and $\bar{p}p$ total cross sections at high energies. 
Also, from a quantum field theory point of view, a vector-pomeron--pion--pion 
vertex cannot exist due to crossing and $C$-parity relations. 
Thus, a vector exchange cannot be used to describe 
the $pp \to pp \pi^{+}\pi^{-}$ reaction~\cite{Lebiedowicz:2016ioh}.
In addition, as was discussed in~\cite{Britzger:2019lvc},
a vector pomeron cannot contribute to the total photoabsorption cross section 
and to the DIS structure functions at low $x_{\rm B}$
in the high-energy diffractive regime. 
All these findings clearly show that the tensor-pomeron 
framework is the only consistent approach.

\item
Comparison of CEP of $\eta$, $\eta'(958)$ at WA102
and at the LHC will allow us to disentangle 
and determine the relative strengths of the leading 
$\Pom\Pom$-fusion contribution and the non-leading exchanges 
involving $f_{2\Reg}$-reggeons, namely 
$\Pom f_{2\Reg} + f_{2\Reg}\Pom$ and 
$f_{2\Reg}f_{2\Reg}$.
We have provided the cross sections expected in LHC experiments, where the $\Pom\Pom$-fusion contribution dominates.
Furthermore, experiments at lower energies, such as HADES and CBM at GSI-FAIR, will also provide valuable insights into the production mechanism of pseudoscalar $\eta$, $\eta'$ and axial-vector $f_{1}(1285)$ mesons.

\item
Finally, we address the argument that the 
$pp \to pp(\Pom \Pom \to \eta)$ reaction should be 
forbidden because the $\eta$ is an SU(3)$_{\rm F}$ octet 
while the pomeron is an SU(3)$_{\rm F}$ singlet. 
This reasoning is invalidated by the significant breaking 
of flavor SU(3)$_{\rm F}$ symmetry and the presence 
of a non-negligible gluon-gluon component in the $\eta$ 
wave function, which enables its direct coupling to pomerons. 
The experimental observation of $\eta$ CEP at the LHC 
will conclusively resolve this issue.

\end{itemize}

%
%

\end{document}